\documentclass[times,twocolumn,final,longtitle]{elsarticle}

\usepackage{framed,multirow}

\usepackage{amssymb}
\usepackage{latexsym}
\usepackage{amsmath}

\usepackage{url}

\usepackage{hyperref}
\usepackage{float}

\usepackage[table]{xcolor}

\usepackage[switch]{lineno}

\usepackage[normalem]{ulem}

\definecolor{deletion}{RGB}{196, 49, 49}
\definecolor{addition}{RGB}{0, 92, 3}

\newcommand{\revised}[2]{{#2}}
\newcommand{\revisednospace}[2]{{#2}}

\journal{Medical Image Analysis}

\begin{document}


\begin{frontmatter}

\title{The state of the art in kidney and kidney tumor segmentation in contrast-enhanced CT imaging: Results of the KiTS19 Challenge}

\author[1]{Nicholas Heller\corref{cor1}}
\cortext[cor1]{Corresponding author: 
  Tel.: +1-612-624-2186;
}
\ead{helle246@umn.edu}

\author[2,3]{Fabian Isensee}
\author[2]{Klaus H. Maier-Hein}
\author[4]{Xiaoshuai Hou}
\author[4]{Chunmei Xie}
\author[4]{Fengyi Li}
\author[4]{Yang Nan}
\author[5,6]{Guangrui Mu}
\author[7]{Zhiyong Lin}
\author[5]{Miofei Han}
\author[5]{Guang Yao}
\author[5]{Yaozong Gao}
\author[8,9]{Yao Zhang}
\author[8,9]{Yixin Wang}
\author[8,9]{Feng Hou}
\author[10]{Jiawei Yang}
\author[10]{Guangwei Xiong}
\author[11]{Jiang Tian}
\author[11]{Cheng Zhong}
\author[12]{Jun Ma}

\author[1]{Jack Rickman}
\author[1]{Joshua Dean}
\author[1]{Bethany Stai}
\author[1]{Resha Tejpaul}
\author[1]{Makinna Oestreich}
\author[1]{Paul Blake}
\author[15]{Heather Kaluzniak}
\author[15]{Shaneabbas Raza}
\author[1]{Joel Rosenberg}
\author[16]{Keenan Moore}
\author[1]{Edward Walczak}
\author[1]{Zachary Rengel}
\author[1]{Zach Edgerton}
\author[1]{Ranveer Vasdev}
\author[1]{Matthew Peterson}
\author[1]{Sean McSweeney}
\author[14]{Sarah Peterson}

\author[13]{Arveen Kalapara}
\author[13]{Niranjan Sathianathen}
\author[1]{Nikolaos Papanikolopoulos}
\author[1]{Christopher Weight}

\address[1]{University of Minnesota, Minneapolis, United States}
\address[2]{German Cancer Research Center (DKFZ), Heidelberg, Germany}
\address[3]{University of Heidelberg, Heidelberg, Germany}
\address[4]{PingAn Technology Co., Ltd, Shanghai, China}
\address[5]{Shanghai United Imaging Intelligence Inc., Shanghai, China}
\address[6]{Southern Medical University, Guangzhou, China}
\address[7]{Peking University First Hospital, Beijing, China}
\address[8]{Institute of Computing Technology, Chinese Academy of Sciences, Beijing, China}
\address[9]{University of Chinese Academy of Sciences, Beijing, China}
\address[10]{Southeast University, Nanjing, China}
\address[11]{AI Lab, Lenovo Research, Beijing, China}
\address[12]{School of Science, Nanjing University of Science and Technology, Nanjing, China} 
\address[13]{University of Melbourne, Melbourne, Australia}
\address[14]{Brigham Young University, Provo, United States}
\address[15]{University of North Dakota, Grand Forks, United States}
\address[16]{Carleton College, Northfield, United States}



\begin{abstract}
There is a large body of literature linking anatomic and geometric characteristics of kidney tumors to perioperative and oncologic outcomes. Semantic segmentation of these tumors and their host kidneys is a promising tool for quantitatively characterizing these lesions, but its adoption is limited due to the manual effort required to produce high-quality 3D segmentations of these structures. Recently, methods based on deep learning have shown excellent results in automatic 3D segmentation, but they require large datasets for training, and there remains little consensus on which methods perform best. The 2019 Kidney and Kidney Tumor Segmentation challenge (KiTS19) was a competition held in conjunction with the 2019 International Conference on Medical Image Computing and Computer Assisted Intervention (MICCAI) which sought to address these issues and stimulate progress on this automatic segmentation problem. A training set of 210 cross sectional CT images with kidney tumors was publicly released with corresponding semantic segmentation masks. 106 teams from five continents used this data to develop automated systems to predict the true segmentation masks on a test set of 90 CT images for which the corresponding ground truth segmentations were kept private. These predictions were scored and ranked according to their average S\o rensen-Dice coefficient between the kidney and tumor across all 90 cases. The winning team achieved a Dice of 0.974 for kidney and 0.851 for tumor, approaching the inter-annotator performance on kidney (0.983) but falling short on tumor (0.923). This challenge has now entered an ``open leaderboard'' phase where it serves as a challenging benchmark in 3D semantic segmentation.
\end{abstract}


\end{frontmatter}


\section{Introduction}
\label{S:intro}

The incidence of kidney tumors is increasing, especially for small, localized tumors that are often discovered incidentally \citep{hollingsworth2006rising}. It is often difficult to radiographically differentiate between benign kidney tumors (e.g., angiomyolipoma and oncocytoma) and malignant Renal Cell Carcinoma (RCC) \citep{millet2011characterization}, but most kidney tumors are eventually found to be malignant \citep{chawla2006natural}. Surgical removal of localized RCC is regarded as curative \citep{capitanio2016renal}, so most localized kidney tumors are removed despite the sizable minority that are postoperatively found to be benign \citep{kim2019association}. 

Traditionally, kidney tumors were removed through \revised{\textit{radical}}{radical} nephrectomy in which the entire kidney along with the tumor are excised \citep{robson1963radical}. However, in order to preserve renal function \citep{scosyrev2014renal}, \revised{\textit{partial}}{partial} nephrectomy, where only the tumor is removed, has recently become the standard of care in an increasing share of tumors with \revised{\textbf{lower surgical complexity}}{lower surgical complexity} \citep{campbell2017renal}. Further, a growing body of literature suggests that a large proportion of renal tumors are \revised{\textit{indolent}}{indolent} \citep{richard2016active,uzosike2018growth,mcintosh2018active,patel2016prospective}, meaning they will never become a danger to the patient, and thus \revised{\textit{active surveillance}}{active surveillance} has emerged as an increasingly popular treatment strategy for tumors exhibiting \revised{\textbf{less aggressive}}{less aggressive} characteristics in imaging. 

With these developments, there is an exciting opportunity to reduce overtreatment of renal tumors without compromising oncologic outcomes \citep{mir2017partial}, but there is a need for methods to objectively quantify the \revised{\textbf{complexity}}{complexity} and \revised{\textbf{aggression}}{aggression} of kidney tumors in order to better inform treatment decisions like radical nephrectomy vs. partial nephrectomy vs. active surveillance. Clinicians predominantly rely on imaging, primarily CT, to assess the complexity and aggression of renal masses. A number of manual scoring systems, termed \revised{\textit{nephrometry}}{nephrometry} scores, have been proposed for this purpose \citep{kutikov2009renal,ficarra2009preoperative,simmons2010kidney}, but they have seen limited adoption due to the significant manual effort they require \citep{simmons2012diameter}, the interobserver variability between expert raters \citep{spaliviero2015interobserver}, and their limited predictive power \citep{kutikov2011anatomic,hayn2011renal,okhunov2011comparison}.

\begin{figure}
    \centering
    \includegraphics[width=\columnwidth]{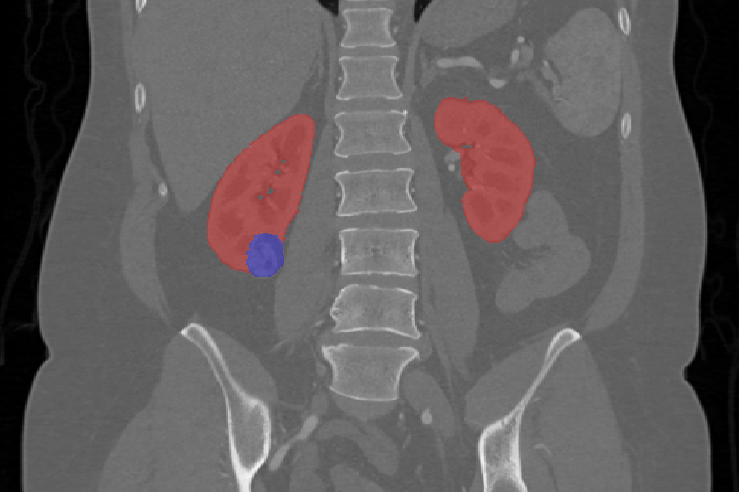}
    \caption{An example of a coronal section of one of the training cases with its ground truth segmentation overlaid (kidney in red, tumor in \revised{green}{blue}). Visualization generated by ITKSnap \citep{yushkevich2016itk}. Best viewed in color.}
    \label{fig:flow}
\end{figure}

Semantic segmentation of kidneys and kidney tumors offers an expressive characterization of the lesion, but it imposes \revisednospace{an }{}an even larger burden of manual effort than most nephrometry scores. Reliable automatic semantic segmentation of kidneys and kidney tumors would enable full automation of several nephrometry scores as well as studies of kidney tumor morphology on unprecedented scales.

The 2019 Kidney and Kidney Tumor Segmentation Challenge (KiTS19) aimed to accelerate progress on this automatic segmentation problem by releasing a dataset of 210 CT images with associated high-quality kidney and kidney tumor segmentations that could be used for training learned models. It also aimed to objectively assess the state of the art by holding a collection of 90 segmentation masks private for participants to predict given associated imaging. Participating teams were ranked based on their average S\o rensen-Dice coefficient between the kidneys and tumors across these 90 test cases. 

This challenge was hosted on grand-challenge.org\footnote{\url{https://kits19.grand-challenge.org}} where it accrued 826 registrations prior to the deadline. 106 unique teams submitted valid predictions to the challenge, and the official leaderboard\footnote{\url{http://results.kits-challenge.org/miccai2019}} reflects submissions from 100 unique teams who met all criteria for a complete submission, including a detailed manuscript describing their method. This challenge was accepted to be held in conjunction with the 2019 International Conference on Medical Image Computing and Computer Assisted Intervention (MICCAI) in Shenzhen, China, and it has now entered an indefinite ``open leaderboard'' phase in which any grand-challenge.org user may submit their predictions on the 90 test cases and have them scored and added to the leaderboard without delay. 

The remainder of this manuscript is structured as follows: In Section \ref{S:rr} we give a brief description of prior work in 3D segmentation as well as existing public datasets and prior challenges. In Section \ref{S:mnm} we describe the design of the challenge, including the dataset, rules, timeline, and evaluation metric. In Section \ref{S:res} we provide an in-depth description of the methods of the top five placing teams, and discuss the implications of these high-performing methods in the context of 3D segmentation research. In Section \ref{S:lims} we discuss the limitations of this challenge, including issues with the dataset, challenge design, and what could be done to address these in future iterations. Finally we give concluding remarks in Section \ref{S:conc}.

\section{Related Work}
\label{S:rr}

\subsection{Biomedical Grand Challenges}
\label{SS:chl}
``Grand Challenges'' in biomedical image analysis are events in which participants compete against one another to develop automated (or sometimes semi-automated) systems that perform a task in medical imaging (i.e., given some input, produce a particular output). Challenge \revised{\textit{organizers}}{organizers} clearly define how participating teams will be evaluated, and usually release some training data to help teams develop \revised{systems based on machine learning}{their prediction models}. The use of common benchmarks such as this have a long history in predictive learning \citep{west1997comparison,maier2018rankings}, but they first started to attract broader interest within the medical imaging community in 2007 when ``Biomedical Grand Challenges'' were first officially affiliated with the MICCAI conference in Brisbane Australia that year \citep{heimann2009comparison}. Since then, hundreds of challenges have been organized in conjunction with  a wide array of related conferences \citep{reinke2018exploit}.

Challenges serve two main purposes in biomedical image analysis research:
\begin{enumerate}
    \item They allow for the \revised{\textbf{objective and fair comparison}}{objective and fair comparison} of methods. With all participating teams evaluated on a single benchmark that's centrally maintained by the organizers, performance metrics are consistent, and the test set performance isn't subject to sampling variations, each of which can obscure comparisons between independent studies on separate, private data.
    \item They contribute \revised{\textbf{publicly available data}}{publicly available data} to the research community. This stimulates and democratizes the development of systems for the target task and sometimes for other related tasks. This work would otherwise be limited to entities with the resources to curate large quantities of high-quality annotated data.
\end{enumerate}

\revisednospace{It`s}{It is} important that new challenges continue to be organized because the set of interesting tasks in biomedical image analysis is vast and largely untapped. Even if each task could be adequately covered, the risk that high-performing methods are overfitting to the quirks of that particular test set rather than the true data generating process grows over time. Further, no challenge is perfect, and organizers continue to learn and improve from the experiences of others. We enumerate the limitations of the present challenge in Section \ref{S:lims} for exactly this reason.

\subsection{3D Segmentation}
\label{SS:3dseg}
3D semantic segmentation is the voxel-wise delineation of regions in three-dimensional imaging such as CT or MRI. Several diverse applications of semantic segmentation have been proposed including targeting for radiation therapy \citep{he2019multi}, patient-specific surgical simulation \citep{taha2018kid}, and quantitative diagnostic and prognostic scoring of anatomic and histologic indications of disease \citep{farjam2007image,tang2018semi,blake2019automatic}. However, without automation of the upstream segmentation task, the prospects for clinical translation of these applications is poor. Automatic semantic segmentation of biomedical imaging has thus arisen as an important and popular research direction. Indeed, roughly 70\% of prior biomedical grand challenges were focused on \revised{}{either instance or} semantic segmentation \citep{maier2018rankings}.

The third spatial dimension presents a unique set of challenges to this field such as a significantly higher cost-per-instance for data annotation and higher memory \revised{intensity}{consumption} during model training and inference. Further, because this problem is relatively esoteric compared with its 2D counterpart, the prospect for leveraging transfer learning \citep{bengio2012deep} from huge, well-established computer vision benchmarks like ImageNet \citep{deng2009imagenet} or MSCOCO \citep{lin2014microsoft} is severely limited. Despite these challenges, recent work in this area has demonstrated impressive performance on the 3D segmentation of a wide variety of anatomical structures and lesions in cross-sectional imaging \citep{maier2017isles,bakas2018identifying,bilic2019liver,zhuang2019evaluation}.

Throughout computer vision, methods based on Deep Learning (DL) now dominate \citep{litjens2017survey} and 3D segmentation is no exception \citep{shen2017deep}. Deep Neural Networks (DNNs) for any given task have a massive design space including a vast array of network architectures, optimization algorithms, and preprocessing procedures. Further, the impressive performance of DL has attracted the many researchers now searching this design space for optimal performance. Unfortunately, with such a high computational cost of training DNNs, most papers that propose a new DNN architecture lack comprehensive benchmarking against the state of the art, especially when results are reported only on private datasets--a practice that still encompasses more than half of papers accepted at MICCAI each year \citep{heller2019role}.

U-Net \citep{ronneberger2015u} and its 3D variants \citep{milletari2016v,cciccek20163d} are some of the earliest proposed methods for DL-based medical image segmentation. In the years since, innumerable modifications have been proposed to U-Net (e.g., residual connections \citep{milletari2016v}, dense connections \citep{li2018h}, and attention mechanisms \citep{oktay2018attention}), with researchers often reporting substantial improvements over their baseline U-Net results. Recently, however, \cite{isensee2018nnu} demonstrate state of the art performance in several well-established 3D segmentation grand challenges using only a U-Net and a novel methodology to search a small space of hyperparameters and preprocessing procedures. This paradigm, termed the nnU-Net, won the recent ``decathalon'' grand challenge\footnote{\url{https://decathlon.grand-challenge.org/}}, in which teams were challenged to develop a system capable of performing well on the tasks of 10 grand challenges simultaneously. 

The authors of nnU-Net make clear that despite its dominance, they do not believe it to be a global optimum of design space, rather they believe it to be a more consistent stepping-off point from which to evaluate architectural ``bells and whistles'' which could present added performance. The KiTS19 challenge is significant because \revisednospace{it`s}{it is} one of the first grand challenges on 3D segmentation to be held after nnU-Net demonstrated its dominant performance and its implementation was released. Thus, one might have expected KiTS19 to be won by the team which finds the right architectural garnish with which to augment the nnU-Net baseline, but as we discuss in Section \ref{S:res}, the ``vanilla'' nnU-Net demonstrated top performance despite attempts by several teams to enhance it.

\section{Materials and Methods}
\label{S:mnm}
\subsection{The KiTS19 Dataset}
We obtained approval from the University of Minnesota Institutional Review Board to conduct a retrospective review of patients who underwent partial or radical nephrectomy for suspicion of renal cancer by a physician in the urology service of University of Minnesota Health between 2010 and mid-2018. 544 patients met this initial criteria. We then excluded patients for whom imaging in the late arterial phase was not available. By the semantic definitions that we chose to use, cyst was regarded as a part of the kidney. We therefore had to exclude \revised{a small number of}{two} patients whose tumor was postoperatively determined to be a cyst. We also excluded patients with tumor thrombus, since in these cases the tumor extends out beyond what we considered to be the primary site and the appropriate boundaries were ambiguous. This left the 300 patients comprising the KiTS19 dataset.

\revisednospace{}{The exclusion of misdiagnosed cysts based on postoperative data presents a dilemma regarding the preoperative use of models trained and validated on this data, since the dataset is not quite a representative sample of all patients who might undergo nephrectomy on suspicion of renal cancer. This sampling bias was weighed against the consequences of introducing clear semantic conflicts in our class labels. We chose to accept the former to avoid the latter, but this is an area where future challenges could improved on KiTS19, for example by explicitly labeling benign cysts as their own class. Since the exclusion of cases with thrombus was determined using preoperative data only, these exclusions (27/329) introduce no further sampling bias.}

\begin{figure}
    \centering
    \includegraphics[width=0.85\columnwidth]{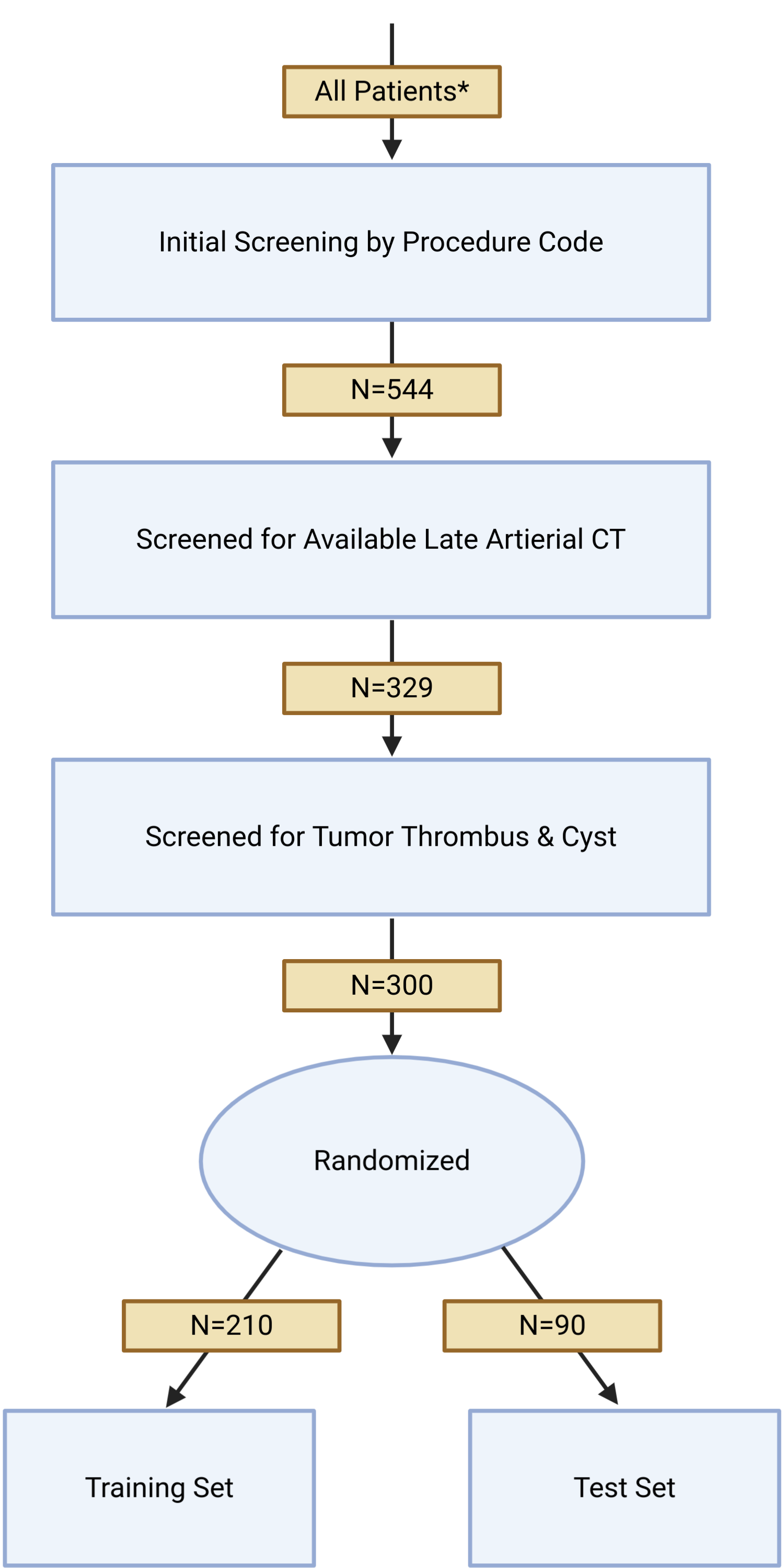}
    \caption{A flow chart of the patient inclusion and random assignment to either training or test set of the study. \revisednospace{}{All Patients* refers only to patients who have indicated that their data may be used for research purposes. This is the default status for patients in Minnesota.}}
    \label{fig:mesh1}
\end{figure}

The imaging for these 300 patients was downloaded in DICOM format and converted to Nifti \citep{larobina2014medical} using the pydicom \citep{mason2011t} and nibabel \citep{brett2018nipy} packages in Python 3.6\footnote{https://www.python.org/}. Students under the supervision of the challenge's clinical chair, Dr. Christopher Weight, then produced manual delineations for the kidneys and tumors in these images using a web-based annotation platform developed in-house \citep{heller2017web}. \revisednospace{}{Each case underwent a single annotation process which involved continuous refinement by a number of annotators in series. This process ended with a final check and refinement by author NH against the radiology and pathology notes. In order to quantify the variability of the process, the entire process--including the review and refinement by NH--was repeated on a randomly selected set of 30 cases from the training set. This was done after the challenge deadline, so these second annotations were not used for the challenge in any way other than to estimate interobserver agreement. A comprehensive description of the annotation protocol and quality assurance procedures can be found in \cite{heller2019kits19}.}

\revisednospace{}{A semantic segmentation paradigm was chosen over instance segmentation due to the inherent ambiguity in the definition of ``instances'' of kidney and tumor. For example, the true border between two abutting tumors is not well-defined, even if they're different subtypes and even at the level of histology. Similarly, in some cases the medial portion of the left and right kidneys are joined just above the spine (termed ``horseshoe kidney,'' representing two of the 300 cases in our cohort), and the boundary between the two is similarly ambiguous. In the authors` opinion, semantic segmentation yields virtually all the same clinical utility of instance segmentation without introducing needless ambiguity in the labels.}

\revisednospace{}{The dataset is available for use in accordance with the terms of the Creative Commons Attribution-NonCommercial-ShareAlike 4.0 International (CC BY-NC-SA 4.0) license. Note that this did not preclude industry-affiliated teams from competing in the challenge, and indeed several companies were represented such as Lenovo, IBM, and NVIDIA.}

Despite the fact that all of these patients were treated by \revisednospace{U}{u}rologists at a single tertiary care center, more than half of the imaging was acquired across more than 50 referring institutions. This, combined with the range of acquisition protocols within each institution, made KiTS19 a diverse dataset in terms of the voxel dimensions, contrast timing, table signature, and scanner field of view. This helps to ameliorate some concern over the external validity of a single institution retrospective cohort like KiTS19, but in future work, a multi-institution cohort with a prospectively collected test set would be preferable \citep{park2018methodologic}. Table \ref{tab:baselines} provides a summary of the patient characteristics.

\begin{figure}
    \centering
    \includegraphics[width=\columnwidth]{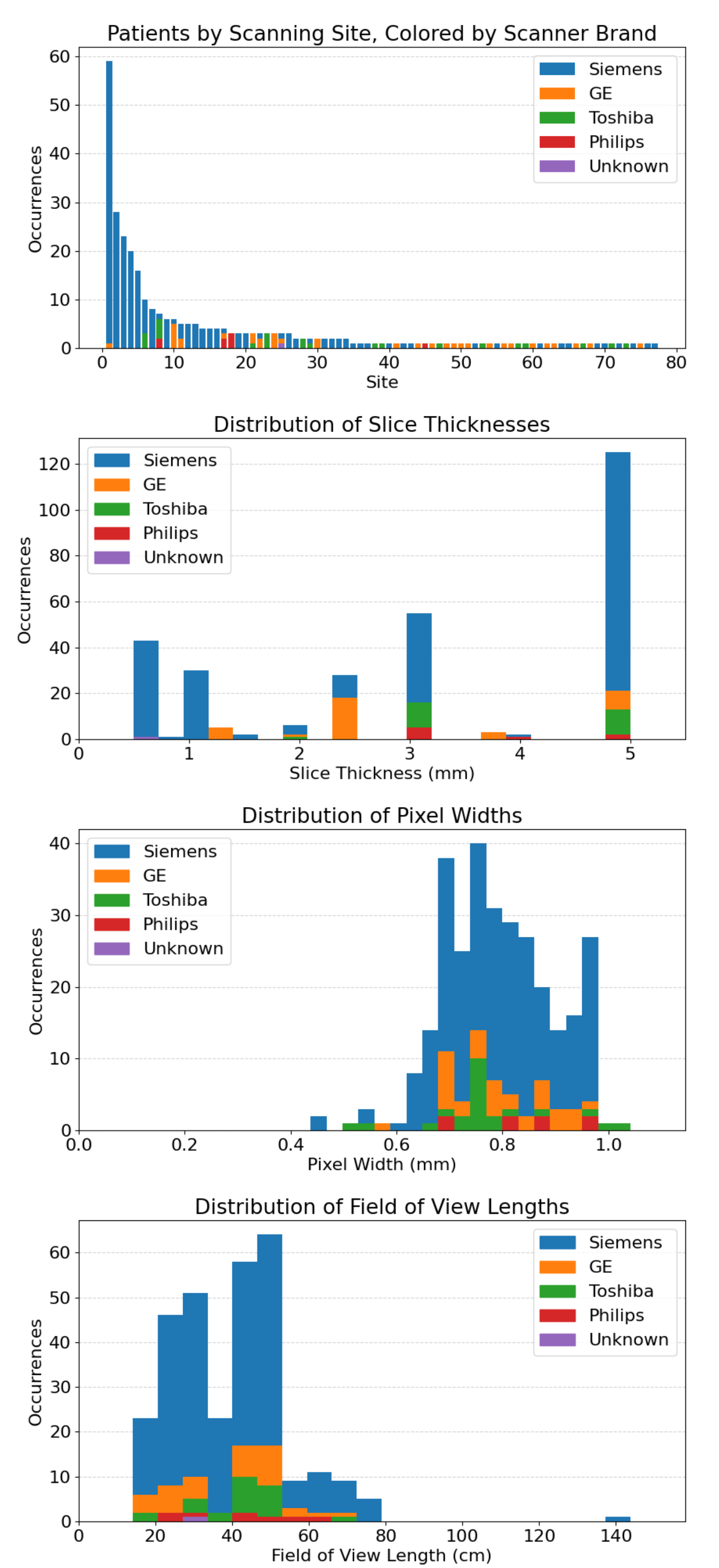}
    \caption{\revisednospace{}{Distributions of scanning sites (top), pixel widths (second), slice thicknesses (third), and longitudinal fields of view (bottom) in the dataset. Scans are colored by scanner manufacturer. Their totals are Siemens: 233, GE: 35, Toshiba: 23, and Philips: 8.}}
    \label{fig:mesh1}
\end{figure}

\renewcommand{\arraystretch}{1.5}
\begin{table}[h!]
{
\begin{tabular}{|p{0.45\columnwidth}|p{0.45\columnwidth}|}
\hline
\rowcolor{gray!70} \textbf{Attribute} & \textbf{Values (N=300)} \\
Age \revisednospace{}{(years)} & 60 (51, 68) \\
\rowcolor{gray!10}BMI \revisednospace{}{(kg/$\text{m}^2$)} & 29.82 (26.16, 35.28) \\
Tumor Diameter* \revisednospace{}{(cm)} & 4.200 (2.600, 6.125) \\
\rowcolor{gray!10}\revisednospace{}{Tumor Volume* ($\text{cm}^3$)} & \revisednospace{}{34.93 (9.587, 109.7)} \\

\rowcolor{gray!40}\multicolumn{2}{|l|}{\textbf{Gender}} \\
\hspace{0.2cm} Male & 180 (60\%) \\
\rowcolor{gray!10}\hspace{0.2cm} Female & 120 (40\%) \\
\rowcolor{gray!40}\multicolumn{2}{|l|}{\textbf{Procedure}} \\
\rowcolor{gray!10}\hspace{0.2cm} Radical Nx & 112 (37.3\%) \\
\hspace{0.2cm} Partial Nx & 188 (62.6\%) \\
\rowcolor{gray!40}\multicolumn{2}{|l|}{\textbf{Subtype}} \\
\hspace{0.2cm} Clear Cell RCC & 203 (67.7\%) \\ 
\rowcolor{gray!10}\hspace{0.2cm} Papillary RCC & 28 (9.3\%) \\ 
\hspace{0.2cm} Chromophobe RCC & 27 (9\%) \\ 
\rowcolor{gray!10}\hspace{0.2cm} Oncocytoma & 16 (5.3\%) \\
\hspace{0.2cm} Other & 26 (8.7\%) \\
\hline
\rowcolor{gray!40}\multicolumn{2}{|l|}{\textbf{\revisednospace{}{Tumor Focality}}} \\
\rowcolor{gray!10}\hspace{0.2cm} \revisednospace{}{Unifocal} & \revisednospace{}{288 (96\%)} \\
\hspace{0.2cm} \revisednospace{}{Unilateral Multifocal} & \revisednospace{}{6 (2\%)} \\
\rowcolor{gray!10}\hspace{0.2cm} \revisednospace{}{Bilateral} & \revisednospace{}{6 (2\%)} \\
\rowcolor{gray!40}\multicolumn{2}{|l|}{\textbf{\revisednospace{}{Renal Anatomy}}} \\
\hspace{0.2cm} \revisednospace{}{Normal} & \revisednospace{}{293 (97.7\%)} \\
\rowcolor{gray!10}\hspace{0.2cm} \revisednospace{}{Solitary} & \revisednospace{}{5 (1.67\%)} \\
\hspace{0.2cm} \revisednospace{}{Horseshoe} & \revisednospace{}{2 (0.67\%)} \\
\hline\end{tabular}
}
\caption{The baseline and tumor characteristics of patients in the KiTS19 dataset. Continuous variables are reported as: Median (Q1, Q3). *In cases where there is more than one tumor, \revised{the largest tumor diameter was reported}{measurement on the largest tumor is reported}. The initialism RCC represents Renal Cell Carcinoma.}
\label{tab:baselines}
\end{table}

\subsection{Design of the KiTS19 Challenge}

In \cite{reinke2018exploit}, the authors review challenges hosted in conjunction with MICCAI conferences from 2007 to 2016 with a focus on how common weaknesses in the design of these challenges made them vulnerable to leaderboard manipulation, which could seriously jeopardize the validity of these challenges as objective and fair benchmarks. We found this paper to be very instructive in the design of this challenge, and we made every effort to follow the best practices outlined by the authors.

\subsubsection{Rules}

\begin{enumerate}
    \item \textbf{External Data:} Teams \revised{\textit{were}}{were} permitted to use external data, as long as that data was \revised{\textit{publicly available}}{publicly available}. This was intended to allow for methods based on pre-training and domain adaptation without allowing teams who might have internal datasets to have an unfair advantage.
    \item \textbf{Replicability:} Teams were required to submit a manuscript describing their methods in detail, as well as any external sources of data. Teams were \revised{\textit{highly encouraged}}{highly encouraged} but not required to make their code available as well.
    \item \textbf{Submissions per Team:} Teams were allowed to enter \revised{\textit{only one}}{only one} submission to the final leaderboard. This was to prevent ``fine tuning'' on the test set by sending many different submissions and simply retracting all lower-scoring entries once the scores are released, which would result in an over-estimate of true performance. This policy was strictly enforced, and several cases in which a team made multiple submissions from different accounts were recognized and only the most recent submission was considered.
    \item \textbf{Manual Predictions:} The predictions submitted to the challenge were required to be entirely automatic; any manual intervention in the inference stage was strictly prohibited. 
\end{enumerate}

\subsubsection{Timeline}
\label{SSS:timeline}

The data was collected and annotated between July and December 2018, after which time the challenge was designed and proposed to MICCAI 2019 via their online submission platform. We were notified that KiTS19 was accepted on March 3, 2019, and we then published the official webpage at \url{http://kits19.grand-challenge.org} on March 4.

The training data was released in full on March 15, at which time participants were invited to inspect and suggest improvements to the data for a period of 20 days. After this period, we addressed several concerns with the metadata as well as a handful of segmentation labels and the data was ``frozen'' for the challenge on April 15. On April 23, a second version of the data was released in which the imaging and segmentation labels were resampled to the median spacing of the original dataset: 3mm x 0.781mm x 0.781mm on the longitudinal, anterior-posterior, and mediolateral axes respectively. This was requested by a number of teams who wanted to work with data in a consistent spacing, but did not feel comfortable resampling the data and labels themselves. 

On July 15, the test imaging was released (without labels) and the submission period was opened until July 29. The MICCAI 2019 leaderboard was then released on July 30, and submission reopened indefinitely for the open leaderboard on August 12, 2019. 

\subsubsection{Infrastructure}
\label{SSS:infrastructure}

\revisednospace{The main challenge webpage as well as submission and evaluation platforms were graciously hosted by grand-challenge.org}{}
\revisednospace{. The data for this challenge was hosted on GitHub}{}
\revisednospace{ using their large file storage solution, Git-LFS. A separate server was then set up to run a Discourse}{}
\revisednospace{ discussion forum. The roles of the three websites (grand-challenge.org, github.com, and discourse.kits-challenge.org) were rules and leaderboard management.  Strict rules were never implemented for which types of correspondences should be submitted as GitHub issues and which should be forum posts, but generally speaking GitHub was used for issues with the data and starter code, and Discourse was used for all other correspondences such as clarifications about the rules and official announcements.}{}

\revisednospace{}{Four web servers were used to manage this challenge:}

\begin{enumerate}
    \item \revisednospace{}{\url{https://kits19.grand-challenge.org/}: This is the official homepage of the challenge. It included participation instructions, official rules, the submission platform, and the open leaderboard.}
    \item \revisednospace{}{\url{https://github.com/neheller/kits19/}: This is the source of the official data for the challenge, as well as some starter code for loading and visualizing the cases. Some discussion of problems and questions regarding the dataset and code has occurred in the ``issues'' section of the repository. The large image files were served using Git-LFS, GitHub's large file storage solution\footnote{https://git-lfs.github.com/}}
    \item \revisednospace{}{\url{https://discourse.kits-challenge.org/}: This is the official discussion forum for the challenge. Topics for discussion here included clarifications of the rules, participants looking for team members, and administrative announcements.}
    \item \revisednospace{}{\url{http://results.kits-challenge.org/miccai2019/}: This is a static HTML document which provides the official leaderboard of teams who participated in the MICCAI-affiliated challenge. The ``open leaderboard'' on the first server continuously updates with new submissions, but the leaderboard here will remain fixed. This webpage was built to be interactive such that when you hover on a submission, all other submissions that are not statistically significantly different are highlighted (alpha = 0.05, see section \ref{ss:statistical} for more details on this). Javascript must be enabled for this feature to function correctly.}
\end{enumerate}

\subsubsection{Submission and Evaluation}

The evaluation metric for this challenge was the average S\o rensen-Dice coefficient between kidney and tumor across all 90 test cases. That is
$$S = \frac{1}{90}\sum_{i=210}^{299}\left(\frac{\left|P^{(i)}_{k}\cap T^{(i)}_{k}\right|}{\left|P^{(i)}_{k}\right|+\left|T^{(i)}_{k}\right|}+\frac{\left|P^{(i)}_{t}\cap T^{(i)}_{t}\right|}{\left|P^{(i)}_{t}\right|+\left|T^{(i)}_{t}\right|}\right)$$

where $S$ is the final score, the superscript $(i)$ represents the $i$th case, $P^{(i)}_{k}$ represents the set of predicted kidney voxels, $T^{(i)}_{k}$ represents the set of ground truth kidney voxels, $P^{(i)}_{t}$ represents the set of predicted tumor voxels, and $T^{(i)}_{t}$ represents the set of ground truth tumor voxels. Here, $i$ ranges from 210 to 299 because by our indexing, cases 0 to 209 comprised the training set. 

During the submission period, teams had the option of viewing the \revised{\textit{approximate}}{approximate} score of two of their submissions. By ``approximate'', we mean that 45 cases were sampled from the test set without replacement and the score was calculated for these 45 cases only. For each submission, teams were contacted via email to confirm that the submission had been received, and asked whether they would like to view this approximate score. A ``yes'' response would be honored no more than twice. This helped to alleviate participants` concerns that the predictions were not packaged or interpreted correctly, without providing participants with enough information for test set fine-tuning.

\begin{figure}
    \centering
    \includegraphics[width=\columnwidth]{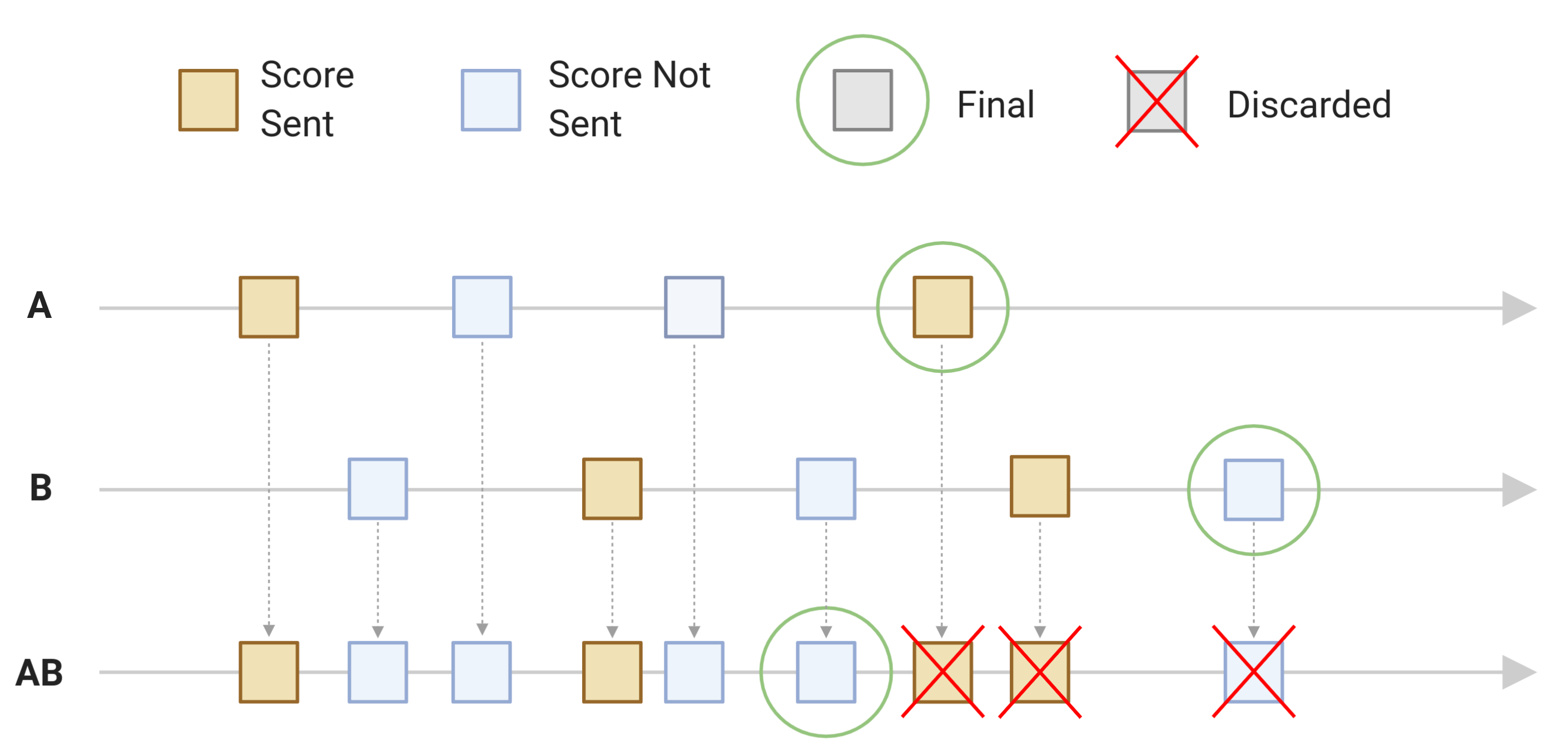}
    \caption{\revisednospace{}{A graphical representation of how sets of submissions are merged if they are later discovered to be made by the same team. "Score Sent" refers to submissions for which the authors requested and received their approximate score.}}
    \label{fig:nogame}
\end{figure}

In an attempt to prevent teams from gaming this system to receive more than two approximate scores \revisednospace{}{by creating sockpuppet accounts}, teams were required to enclose a manuscript with at least one page of content and their final authors list with all submissions. Authors lists were cross-checked against prior submissions before scores were released, and in a few cases where substantially overlapping submissions went unnoticed, all submissions after the one immediately following receipt of the second score were discarded. The final score on the MICCAI leaderboard reflects the most recent remaining submission from each team \revisednospace{}{excluding discarded submissions}, whether its approximate score was requested or not. \revisednospace{}{For example, suppose a team submits four times under username `A`, and asks for approximate scores for submissions one and four, which were honored. At this point, submission four would be considered for the final leaderboard. Suppose that during the audit period after submissions have closed, a second team under username `B` is identified to be the same team as `A`, and has made five submissions, requesting approximate scores for 2 and 4. In this case, approximate scores computed for A and B will be ordered chronologically according to when they were sent, and the most recent submission before the third approximate score is sent to A or B will be considered, while all others are discarded. Fig \ref{fig:nogame} shows this graphically.}

\section{Results}
\label{S:res}
126 unique users on grand-challenge.org made a submission to this challenge, but 20 of these were determined to belong to a team that had submitted previously. A further 6 submissions were disqualified for not providing a sufficient manuscript, even after repeated warnings. This left the 100 teams that appear on the official MICCAI leaderboard. 

Submissions to this challenge were \revised{overwhelmingly}{entirely} based on deep neural networks, although they varied considerably in decisions around preprocessing strategies, architectural details, and training procedures. Sections \ref{SSS:P1} - \ref{SSS:P5} outline the methods used by the five highest-scoring teams listed on the official MICCAI leaderboard. These five teams were also invited to present this material orally at the KiTS19 satellite event of MICCAI 2019 in Shenzhen, China on Oct. 13, 2019.

\subsection{Analysis of Segmentation Performance}
\label{ss:statistical}



\revisednospace{}{Teams had an average tumor Dice of 0.580 (SD: 0.212) which was much lower than the mean kidney Dice of 0.915 (SD: 0.0469). This is not surprising since kidneys tend to occupy a much more predictable location with well-defined relationships to surrounding anatomy. Kidneys also have defining features, such as the corticomedullary differentiation and oblong shape with the major axis approximately parallel with the craniocaudal axis. Renal tumors, on the other hand, can appear in all different shapes, sizes, and locations on the kidney. They also lack any pronounced defining characteristics. In general, methods exhibited a higher precision than recall when segmenting the tumor (mean precision 0.658 vs. mean recall 0.586, p=$7.21\times10^{-7}$, which suggests that DNNs often have a difficult time finding the whole tumor, even if they've identified the correct lesion.}

\subsubsection{Predictors of Segmentation Performance}

\revisednospace{}{Given this set of 100 competitive predictions on each of the 90 test cases, we can conduct a quantitative analysis of case characteristics that seem to associate with predictive performance of deep learning systems generally. We hypothesized that the following characteristics might associate with tumor dice values:}
\begin{itemize}
    \item \revisednospace{}{\textbf{Slice Thickness}: Models might have a harder time making sense of the 3D nature of the data in cases where slices are further apart.}
    \item \revisednospace{}{\textbf{Tumor Focality}: Due to the rarity of multifocality, submissions may exhibit a bias towards predicting only a single tumor region (whether implicit in the model or explicit in post-processing).}
    \item \revisednospace{}{\textbf{Longitudinal Length of the Scan's Field of View (FOV Length)}: A longer field of view generally results in greater range of anatomical characteristics being shown, in some cases including everything from the top of the head to the bottom of the feet. Models trained on data that overrepresents abdomical scans only might have difficulty with false positives in these unfamiliar regions.}
    \item \revisednospace{}{\textbf{Tumor Size}: Smaller tumors have a higher proportion of their voxels on the region's border. Since a vast majority of segmentation errors occur on region boundaries, they may exhibit lower dice scores in general. Further, smaller tumors are often easier to miss entirely.}
    \item \revisednospace{}{\textbf{Tumor Subtype}: Renal tumor subtypes are known to have different prototypical radiographic appearances, and certain subtypes might be more recognizeable than others. Certain subtypes (angiomyolipoma, for example) typically exhibits low attenuation which could make them difficult to distinguish from renal cysts.}
\end{itemize}

\revisednospace{}{Results of this analysis can be found in Table \ref{tab:hidden_strata}}\revisednospace{}{. Regression analysis was performed using the R Language and Environment for Statistical Computing \citep{rlanguage}.}

\begin{table}[H]
{
\begin{tabular}{|p{0.35\columnwidth}|p{0.18\columnwidth}|p{0.30\columnwidth}|}
\hline
\rowcolor{gray!70} \textbf{Variable} & \textbf{Coefficient} & \textbf{P-value} \\
Slice Thickness (mm) & $0.0054$ & $0.61$ \\
\rowcolor{gray!10}Multifocal & $-0.0038$ & $0.97$ \\
FOV Length (cm) & $1.4\times10^{-4}$ & $0.90$ \\
\rowcolor{gray!10}Tumor Size (cm) & $0.043$ & $4.8\times10^{-11}$* \\
Clear Cell & $-0.073$ & $0.25$ \\
\rowcolor{gray!10}Papillary & $-0.0044$ & $0.96$ \\
Other Subtype & $-0.13$ & $0.078$ \\
\rowcolor{gray!10}(Intercept) & 0.43 & $3.6\times10^{-7}$* \\
\hline\end{tabular}
}
\caption{\revisednospace{}{Multivariate regression against average tumor dice across all teams for each case. Statistically significant p values are marked with asterisks.}}
\label{tab:hidden_strata}
\end{table}

\subsubsection{Sensitivity to Choice of Metric}

\revisednospace{}{It is difficult to say whether false positives or false negatives are more important in this context. Our primary intended clinical application is segmentation-based characterization (e.g. size, enhancement patterns, etc), but these segmentations could potentially be used for treatment planning (e.g. patient-specific surgical simulation) or for creating intraoperative overlays on the robotic camera view. In the latter two cases, one might argue that false negatives are more serious since they could result in the tumor being only partially removed, but tumor tissue is usually recognizably different from healthy kidney, so this is not an acute concern to surgeons. On the contrary, others might argue that false positives are actually more serious since they could contribute to what many believe is already a significant problem of overly aggressive treatment of renal tumors. Of course we can only loosely speculate about such things, and prospective studies would be necessary in order to observe the true clinical impact of these sorts of errors.}

\revisednospace{}{In the KiTS19 challenge, we chose to rank teams according to their S\o renson Dice score, which is also a special case of the $F_\beta$-Score where $\beta=1$.}

$$F_\beta = \frac{(1+\beta^2)\cdot\text{TP}}{(1+\beta^2)\cdot\text{TP} + \beta^2\cdot\text{FN} + \text{FP}}$$

\revisednospace{}{Where TP, FP, and FN stand for true positives, false positives, and false negatives, respectively. We chose Dice because it’s the simplest and most commonly used metric in cases where the relative importance of precision and recall is equivocal.}

\begin{figure}
    \centering
    \includegraphics[width=\columnwidth]{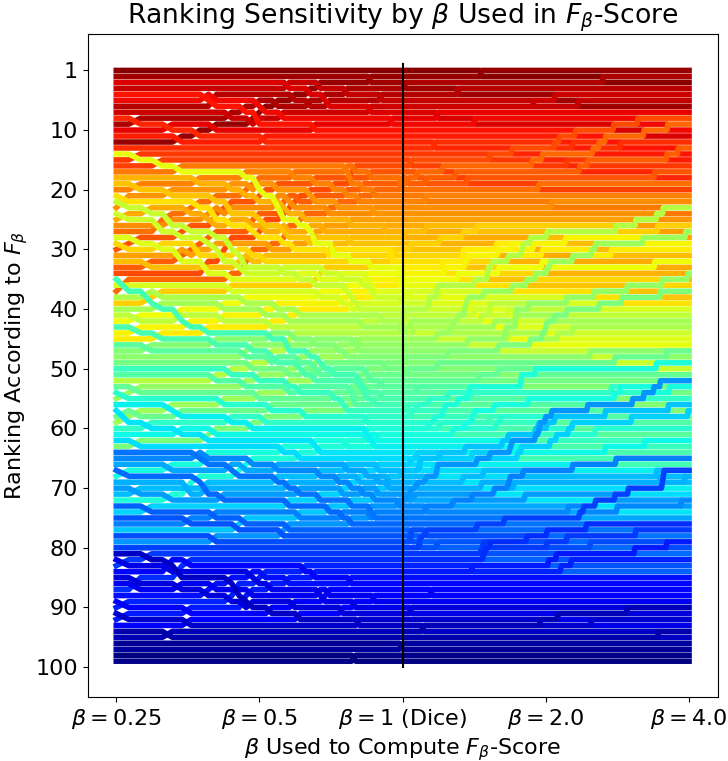}
    \caption{\revisednospace{}{Each submission represents a single line on this plot. Teams are colored according to their official ranking, which explains why a vertical path up the center of the plot (where $\beta=1$) reflects an entirely consistent color progression. Teams which tended to have higher precision but lower recall rank higher with lower $\beta$ values, where teams with higher recall but lower precision tend to rank higher with a higher value of $\beta$. Note that the x-axis is log-scaled.}}
    \label{fig:beta_sensitivity}
\end{figure}

\revisednospace{}{\cite{wiesenfarth2019methods} propose to use a line plot to explore the ranking order’s sensitivity to the chosen metric. Fig. \ref{fig:beta_sensitivity} uses this approach to visualize the effect that the value of $\beta$ has on the ranking order. Stability around beta=1 (also known as the Dice score) appears excellent with even the most extreme cases not traversing more than 15-20\% of the leaderboard. Interestingly, the top and bottom of the leaderboard appear to show the highest stability.}

\subsubsection{Statistical Analysis of Ranking Order}

\begin{figure}
    \centering
    \includegraphics[width=\columnwidth]{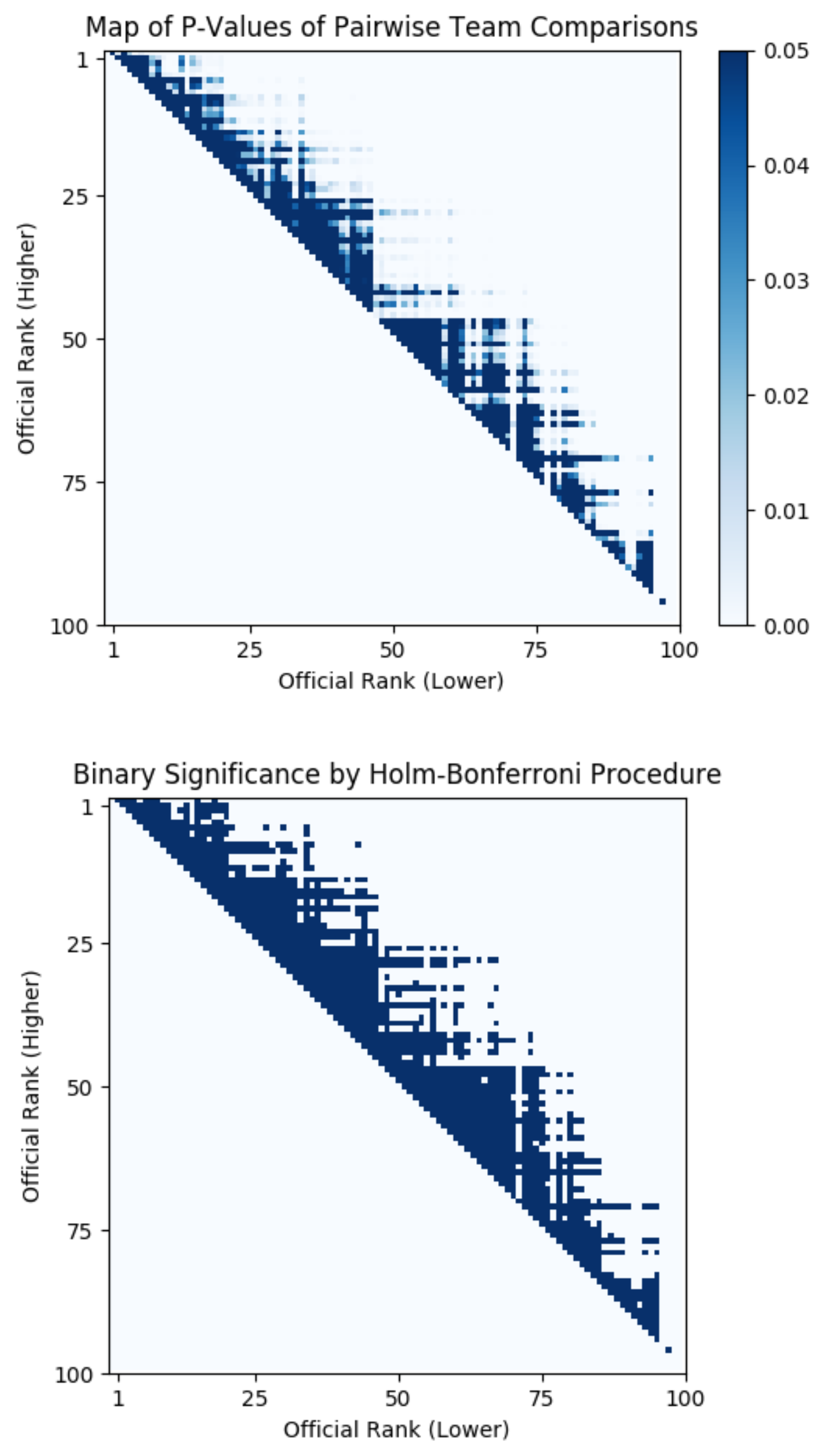}
    \caption{\revisednospace{}{Maps of p-values (top) and statistical significance according to the Holm-Bonferroni procedure (bottom) which corrects for multiple testing. P-values were determined via The Bootstrap \citep{efron1994introduction}. Family-wise error rate ($\alpha$) is set to 0.05, and p-values saturate at 0.05 in the top figure. }}
    \label{fig:sigmap}
\end{figure}

\revisednospace{}{The ranking order of a challenge with many submissions such as KiTS is virtually certain to have some ranking order instability, especially near the median submission where performance is highly concentrated in our case. This can be visualized using a significance map \citep{wiesenfarth2019methods}, as shown in Fig. \ref{fig:sigmap}. In the top image of Fig. \ref{fig:sigmap}, the value of each pixel can be interpreted as the likelihood that the team on the x-axis is actually superior to the team on the y-axis, where the latter holds a higher position on the leaderboard. In the bottom image, pixels are dark if they represent pairs for whom we cannot call the p-values above statistically significant while preserving a 5\% chance of any significant difference representing a false positive.}

\section{Top Five Submissions}

\begin{figure}
    \centering
    \includegraphics[width=\columnwidth]{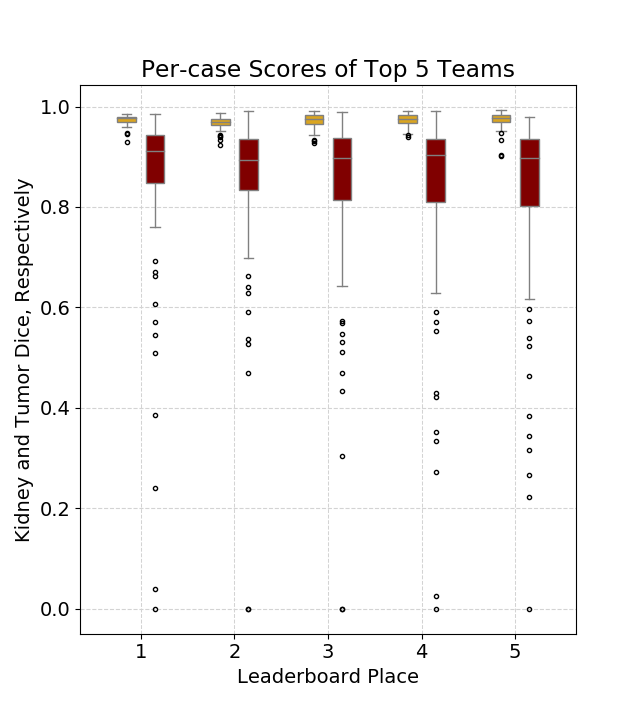}
    \caption{Kidney Dice scores (left, gold) and tumor Dice scores (right, maroon) on each case for each of the top 5 teams. Similar plots for the remaining submissions are included in the supplementary material. \revisednospace{}{A single case was missed (dice < 0.05) by each of the top five teams and by a vast majority of the other 95. In this case, the affected kidney had a concomitant cyst that teams consistently mistook for the tumor.}}
    \label{fig:mesh1}
\end{figure}

\subsection{First Place: An Attempt at Beating the 3D U-Net}
\label{SSS:P1}

This submission was made by Fabian Isensee and Klaus H. Maier-Hein of the German Cancer Research Center. 

\subsubsection{Data Use and Preprocessing}

This submission did not make use of any data other than the official training set. Model parameters were initialized randomly and no transfer learning was used.

The data was downloaded in its original spacing (from the master branch) and resampled to a common spacing of $3.22 \times 1.62 \times 1.62$ mm resulting in median volume dimensions of $128 \times 248 \times 248$ voxels. The CT intensities (HU) were clipped to a range of [-79, 304] and transformed by subtracting 101 and dividing by 76.9. \revisednospace{}{These preprocessing constants were selected through a meta-learning process described in \cite{isensee2019automated}.}

\subsubsection{Architecture}

Three 3D U-Net architectures were tested using five-fold cross-validation. The networks all used 3D convolutions, leaky ReLU (LReLU) activations, and instance normalization. Upsampling was performed with transposed convolutions and downsampling was performed with strided convolutions. The first level of the U-Net extracts either 24 or 30 feature maps, and each downsampling doubles this up to a maximum of 320. Downsampling is stopped once a further downsampling would result in at least one spatial dimension of $<4$ voxels, at which point upsampling begins. The three networks differed as follows:

\textit{Plain 3D U-Net:} This network used 30 feature maps at the highest resolution. Between each up or downsampling, two blocks of conv-instnorm-LReLU were performed.

\textit{Residual 3D U-Net:} This network used 24 feature maps at the highest resolution. In the encoder portion, the conv-instnorm-LReLU were replaced with residual blocks of the form: conv-instnorm-ReLU-conv-instnorm-ReLU where the residual addition takes place before the final activation, similar to \cite{he2016deep}. Just one of these blocks is used at the highest resolution, and with each downsampling another is added. The decoder portion uses just one conv-instnorm-ReLU between upsamplings.

\textit{Pre-activation Residual 3D U-Net:} This network was similar to the Residual 3D U-Net but the residual blocks used \textit{pre-activation} \citep{he2016identity}. The residual blocks were thus instnorm-ReLU-conv-instnorm-ReLU-conv. 

\begin{figure}[H]
    \centering
    \includegraphics[width=\columnwidth]{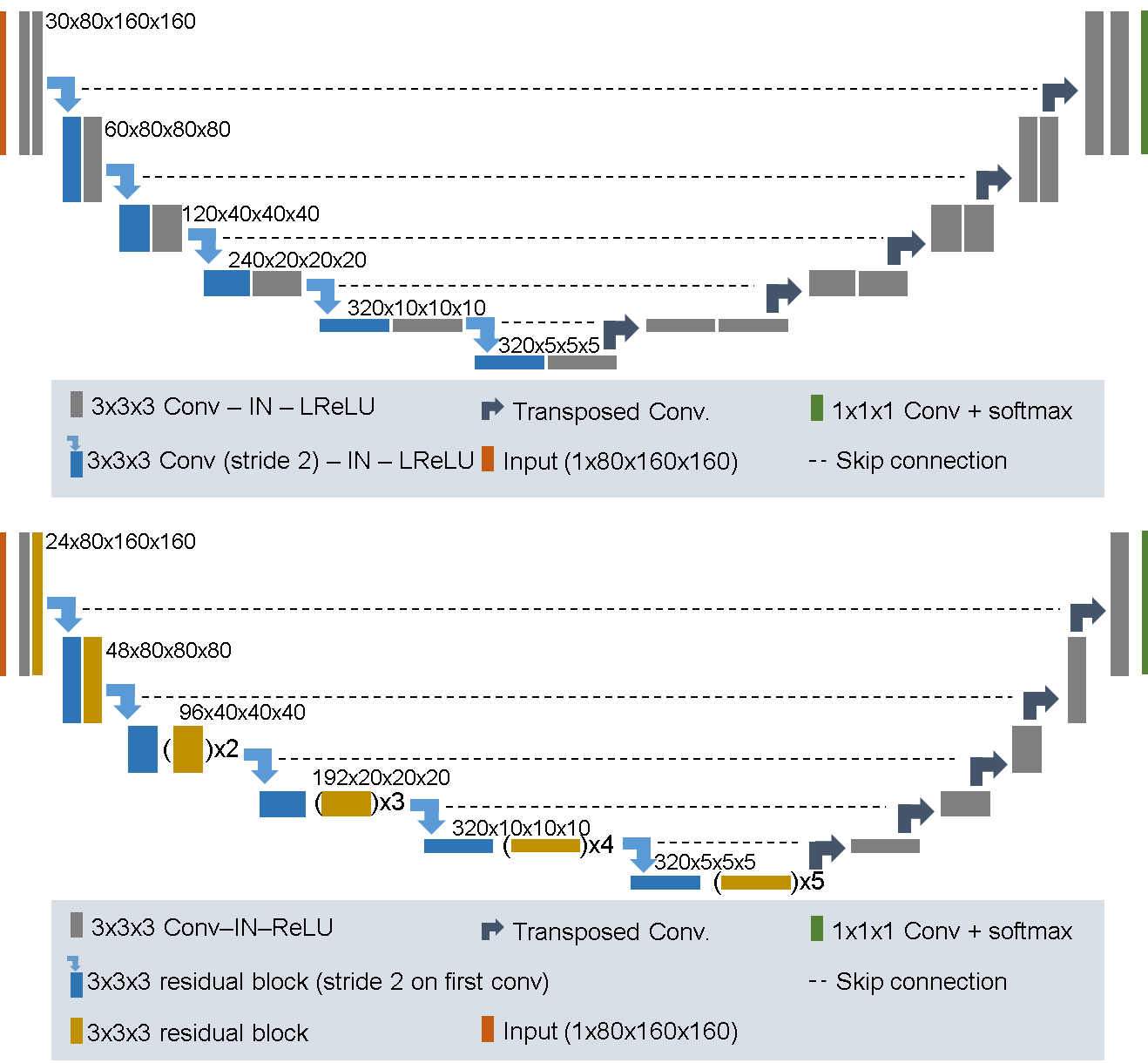}
    \caption{3D U-Net (top) and residual 3D U-Net architecture (bottom) used in this project. $()\times X$ denotes that a block is repeated X times. The architecture of the pre-activation residual U-Net is analogous to the residual U-Net (with instnorm and ReLU being shifted to accommodate pre-activation residual blocks).}
    \label{fig:1st}
\end{figure}

An overview of these architectures is given in Fig. \ref{fig:1st}

\subsubsection{Training}

Patches of size $80 \times 160 \times 160$ were randomly sampled from the resampled volumes for training. 1000 epochs training were performed with an epoch defined as 250 batches with a batch size of two. A sum of cross-entropy and dice loss was used as a training objective with deep supervision. All networks were trained with stochastic gradient descent. Extensive data augmentation with the \textit{batchgenerators} framework\footnote{https://github.com/MIC-DKFZ/batchgenerators} was used during training. Adjustments included scaling, rotations, brightness, contrast, a gamma transformation, and the introduction of Gaussian noise.

Training for each network was performed on a single NVIDIA Titan Xp GPU using the PyTorch framework \citep{paszke2017pytorch} based on the nnU-Net implementation\footnote{https://github.com/MIC-DKFZ/nnUNet} \citep{isensee2019nnu}. Each network took about five days to train. 

The cases that were known to have been mislabeled (outlined in Section \ref{SS:ds}) were excluded from training, and cases 23, 68, 125, 133 were found to be in consistent disagreement with predictions so they were excluded as well. 

Five-fold cross-validation Dice scores of 0.974 and 0.857 were observed for kidney and tumor respectively for the \textit{Residual 3D U-Net} architecture, which was marginally higher than the performance of any of the other approach, including an ensemble of all three \revisednospace{}{architectures}. Therefore, an ensemble of the \revisednospace{}{\textit{Residual 3D U-Net}} networks from \revisednospace{these}{all} five folds only was used for the final test set predictions.

\subsubsection{Postprocessing}

The model predictions were resampled to their original spacing and submitted without any further processing.

\subsubsection{Results}

This submission scored a 0.974 kidney Dice and a 0.851 tumor Dice resulting in a first place 0.912 composite score. For a more detailed description of this submission, see \cite{kits19firstplace}.

\subsection{Second Place: Cascaded Semantic Segmentation for Kidney and Tumor}
\label{SSS:P2}
This submission was made by Xiaoshuai Hou, Chunmei Xie, Fengyi Li, and Yang Nan of PingAn Technology Co. 

\subsubsection{Data Use and Preprocessing}

This submission did not make use of any data other than the official training set. Model parameters were initialized randomly and no transfer learning was used.

The CT intensities were normalized to zero mean and unit standard deviation without clipping. This method had two input modes: it first performed a coarse localization of the kidneys from a low-resolution image (volumes resampled to $1.72 \times 1.72 \times 3.41$ mm), and then performed fine-grained delineation of those kidneys as well as the lesion(s) from a cropped region of high-resolution images (volumes resampled to $0.781 \times 0.781 \times 0.781$ mm). 

\subsubsection{Architecture}

This cascaded approach had three stages. Stage 1 performed a coarse segmentation of all kidneys in the image in order to crop out spatially distant regions for the next stage. This stage is based on the nnU-Net \citep{isensee2018nnu}, but the segmentation masks here are used only to localize the kidney regions, and the predicted masks are discarded. The second stage is run for each rectangular kidney region that is found by the first stage. Here, another 3D U-Net based on the nnU-Net implementation is used to produce a fine-grained segmentation of the kidneys vs background, where tumor is included in the kidney label. Finally, in the third stage of the model, all voxels predicted to be background are set to zero, and a fully convolutional net is used to segment the tumor voxels from the kidney voxels. Here, all predictions made outside of the stage 2 kidney predictions are discarded.

\begin{figure}[H]
    \centering
    \includegraphics[width=\columnwidth]{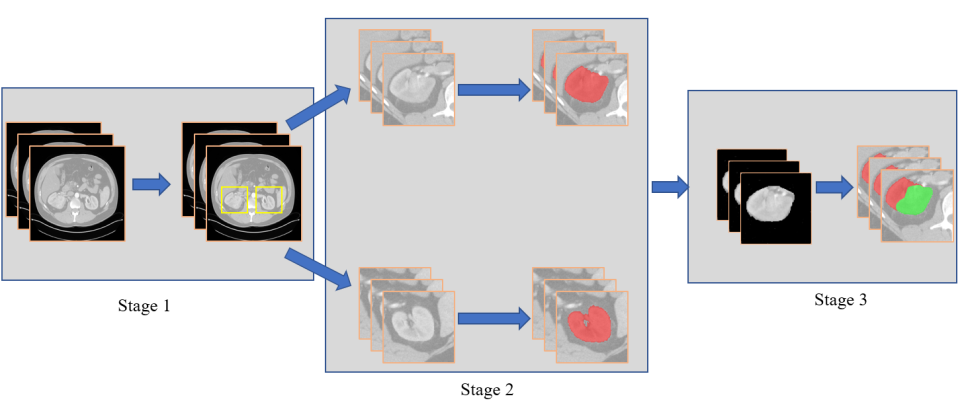}
    \caption{The segmentation pipeline for the second place method. Kidneys are first coarsely localized via segmentation in stage one, then each kidney is finely segmented from background in stage two, and finally the tumor segmented from kidney in stage 3.}
    \label{fig:2nd}
\end{figure}

An overview of this prediction pipeline is shown in Fig. \ref{fig:2nd}.

\subsubsection{Training}

All models were trained with a combination of the cross entropy loss and Dice loss. Data augmentation was used including elastic deformation, rotation, and random cropping. The Adam optimizer \revised{\citep{kingma2014adamkingma2014adam}}{\citep{kingma2014adam}} was used with an initial learning rate of $1e-4$. Whenever the exponential moving average of training loss did not improve by a certain threshold in 30 epochs, the learning rate was scaled by 1/5. Training was terminated if the validation loss did not improve within 50 consecutive epochs. 


\subsubsection{Postprocessing}

After the model made its kidney and tumor predictions, an algorithm was used to fill holes within the tumor prediction and remove some predicted regions that appeared to be false positives.

\subsubsection{Results}

This submission scored a 0.967 kidney Dice and a 0.845 tumor Dice resulting in a second place 0.906 composite score. For a more detailed description of this submission, see \cite{kits19secondplace}.

\subsection{Third Place: Segmentation of kidney tumor by multi-resolution VB-nets}
\label{SSS:P3}
This submission was made by Guangrui Mu, Zhiyong Lin, Miaofei Han, Guang Yao, and Yaozong Gao of Shanghai United Imaging Intelligence Inc. 

\subsubsection{Data Use and Preprocessing}

This submission did not make use of any data other than the official training set. Model parameters were initialized randomly and no transfer learning was used.

The data was downloaded in its original spacing and 30 cases were randomly selected to form a validation set, leaving 180 for training. CT intensity values were clipped to fall in the range of [-200, 500] HU and then uniformly normalized to [-1,1], and all volumes were resampled to an isotropic spatial resolution. A professional doctor then manually delineated all cysts in the dataset in order to help the learned model form an understanding of cyst as well as tumor, and mitigate the risk that the two are confused. 

\subsubsection{Architecture}

The authors of this submission extended the V-Net proposed in \cite{milletari2016v} to include bottlenecks instead of traditional convolutional layers. Here, each bottleneck consists of three convolutional layers -- the first applies a $1 \times 1 \times 1$ kernel and reduces the number of feature maps, the second performs a spatial convolution with some receptive field greater than $1$ in each spatial dimension. The last applies another $1 \times 1 \times 1$ filter to increase the number of feature maps back to their original count. The authors extracted 16 feature maps at the highest resolution, and doubled this with each downsampling. Residual blocks were used throughout the network of the form conv-batchnorm-ReLU-conv-batchnorm-addition-ReLU. Zero padding was used to keep blocks the same size for concatenation. 

This submission also made use of a cascaded approach in which segmentation-based localization was again used on low resolution volumes (voxel size of $6 \times 6 \times 6$ mm) to produce Volumes of Interest (VOIs) which were then fed at a high-resolution (voxel size of $1 \times 1 \times 1$ mm) into a second model which produced final segmentation predictions. 

\begin{figure}[H]
    \centering
    \includegraphics[width=\columnwidth]{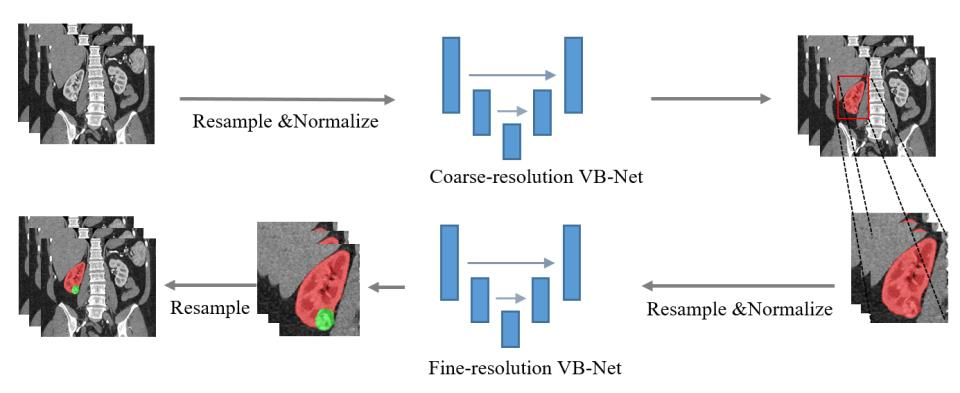}
    \caption{A graphic showing the prediction pipeline for the third place submission to the challenge. Data is first resampled to a low-resolution global image, and kidney segmentation is used to define Volumes of Interest (VOIs). Those VOIs are then resampled to a high resolution, and another network is used to make kidney and tumor segmentation predictions.}
    \label{fig:3rd}
\end{figure}

A graphical representation of this pipeline is shown in Fig. \ref{fig:3rd}.

\subsubsection{Training}

The models were implemented and trained in the PyTorch framework. During the training of both the coarse and fine models, random patches of size $96 \times 96 \times 96$ voxels are sampled from the target volume. A generalized Dice loss is used with the following formulation: 

$$
L = - \frac{1}{C}\sum_{c=1}^{C}\frac{2\sum_i^Np_c(i)g_c(i)}{\sum_i^Np_c^2(i)+\sum_i^Ng_c^2(i)}
$$

Where $C$ represents the number of class labels, $p_c(i)$ is the probability of the class $c$ at voxel $i$ predicted by the network, $g_c(i)\in \{0,1\}$ is the binary label indicating whether the label of voxel $i$ is class $c$.

The Adam \citep{kingma2014adamkingma2014adam} optimization algorithm was used with a constant learning rate of $1e-4$ and a batch size of 6. The networks were trained for 5000 epochs of 30 batches. Data augmentation was not employed. 

\subsubsection{Postprocessing}

Connected component analysis was used to filter out small regions that were predicted as kidney. In order to aid in the discrimination between tumors and cysts with uneven densities, an algorithm made use of the spatial relationship between the cyst and the tumor and their average HU value to give a final, consistent classification. 

\subsubsection{Results}

This submission scored a 0.973 kidney Dice and a 0.832 tumor Dice resulting in a third place 0.903 composite score. For a more detailed description of this submission, see \cite{kits19thirdplace}.

\subsection{Fourth Place: Cascaded Volumetric Convolutional Network for Kidney Tumor Segmentation from CT volumes}
\label{SSS:P4}

This submission was made by Yao Zhang, Yixin Wang, Feng Hou, Jiawei Yang, Guangwei Xiong, Jiang Tian, and Cheng Zhong of the Lenovo AI Lab.


\subsubsection{Data Use and Preprocessing}

This submission did not make use of any data other than the official training set. Model parameters were initialized randomly and no transfer learning was used.

The data from the interpolated branch of the GitHub repository was used for this submission. After download, 50 cases were selected as a validation set, leaving 160 for training. This method also used a course-to-fine approach with a first segmentation-based localization to crop volumes of interest for each kidney in a lower resolution image (voxel spacing $3 \times 1.56 \times 1.56$ mm). Then, those crops were applied to the full resolution data, creating patches that were then sent to a finer-grained segmentation network which made the model`s predictions. Normalization of the CT intensities was done on a case-by-case basis by first clipping values at their 0.5\% and 99.5\% percentiles and then subtracting the mean dividing by the standard deviation. 

\subsubsection{Architecture}

The architecture used for both the coarse localization and fine predictions of this submission is a 3D U-Net based on the nnU-Net implementation. Several hyperparameters and architectural enhancements and their combinations were tested on the validation set and the highest-performing combination was chosen for predictions on the test set. The baseline model that the authors used was a 3D U-Net with instance normalization. The net extracted 30 feature maps at original resolution and doubled this with each downsampling. Downsampling was performed with max-pooling and transposed convolutions were used for upsampling. These up and downsampling operations were strided differently based on the original patch resolution in order to handle blocks with differing extents in each axis. The network downsampled until each spatial dimension of the feature map is smaller than 8 voxels. Leaky ReLU was used for all activation functions outside of the loss layers.

\begin{figure}[H]
    \centering
    \includegraphics[width=\columnwidth]{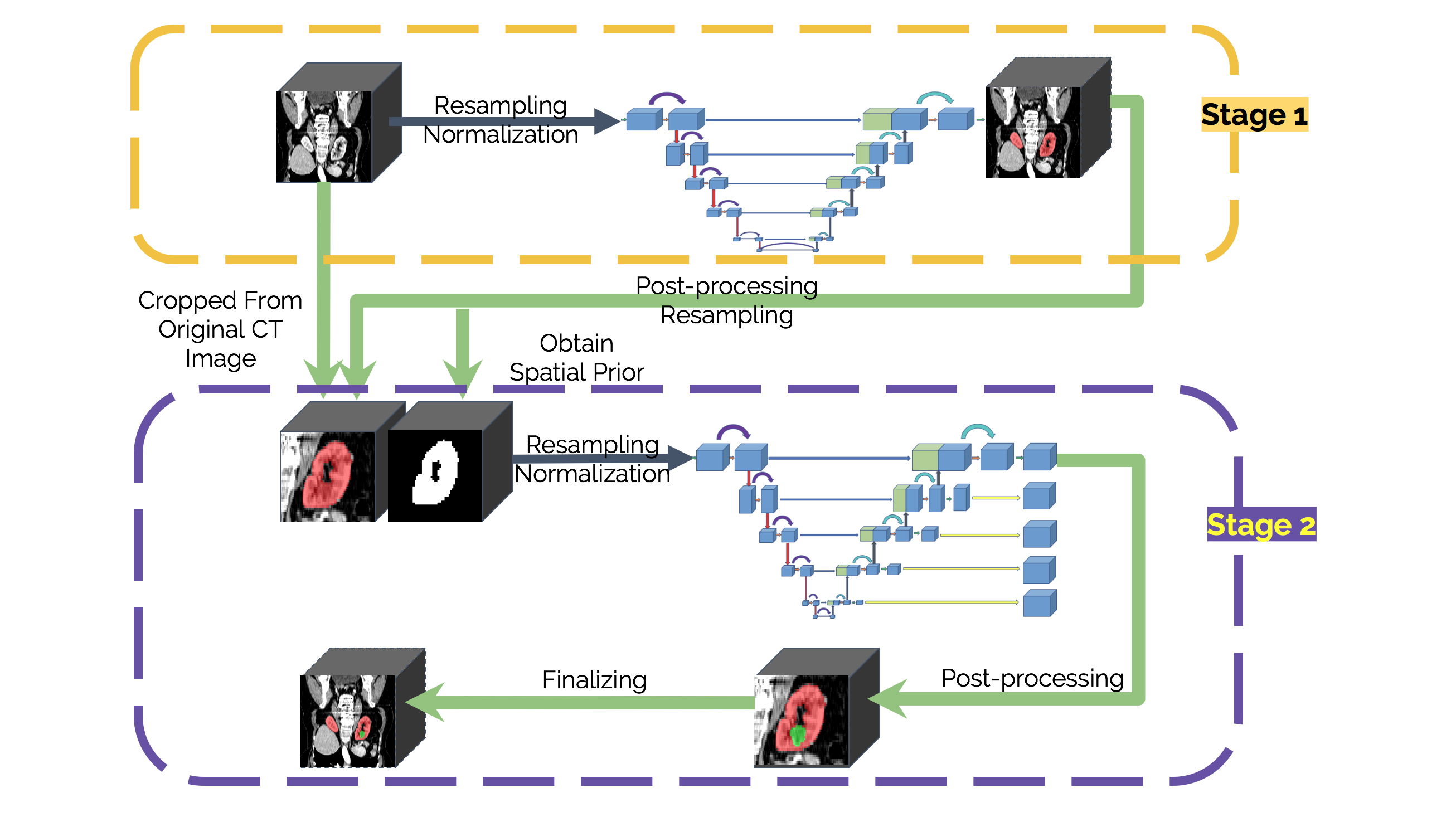}
    \caption{The pipeline used by the fourth place submission. In the first stage, a baseline 3D U-Net equipped with instance normalization and residual blocks is used to obtain the coarse location of the kidney. In the second stage, a counterpart further augmented with deep supervision is employed for both kidney and tumor segmentation.}
    \label{fig:4th}
\end{figure}

The authors tested this baseline against other versions making use of deep supervision, residual blocks between up and down sampling, and the use of a ``spatial prior'', in which they fed the coarse predictions of the first network as an input channel to the second. The authors found that the combination of all three of these modifications yielded the best validation performance, and they thus chose to use this model at test time. A diagram outlining this architecture is shown in Fig. \ref{fig:4th}.

In addition to deep supervision, the authors also employed \revised{\textit{deep prediction}}{deep prediction}. The predictions at each decoder stage of the network are ensembled together by majority voting to produce final predictions.

\subsubsection{Training}

The authors apply data augmentation to expand training data and avoid overfitting. The data augmentation including random rotation, scaling and elastic deformation is implemented on the fly. Random patches of size $40 \times 128 \times 128$ were sampled from the images and used for training with a batch size of 5. The objective was a sum of cross entropy and dice loss at each supervised layer. The Adam algorithm was used for optimization with an initial learning rate of $3e-4$. On a single NVIDIA Tesla V100 32GB GPU, the first stage took roughly 18 hours to train and the second stage took a further 30.

\subsubsection{Postprocessing}

The model operates on the assumption that no more than two kidneys exist in each case. A connected component analysis was run on each case`s prediction to remove all but the two largest components of the kidney`s predicted segmentation.

\subsubsection{Results}

This submission scored a 0.974 kidney Dice and a 0.831 tumor Dice resulting in a fourth place 0.902 composite score. For a more detailed description of this submission, see \cite{kits19fourthplace}.

\subsection{Fifth Place: Cascaded U-Net Ensembles}
\label{SSS:P5}
This submission was made by Jun Ma of the Nanjing University of Science and Technology. 

\subsubsection{Data Use and Preprocessing}

This submission did not make use of any data other than the official training set. Model parameters were initialized randomly and no transfer learning was used.

This submission is built on the data from the interpolated branch of the GitHub repository, and the data were randomly assigned to five folds for cross-validation. The CT intensities were clipped based on the 0.5 and 99.5th percentile and normalized by subtracting the mean and dividing by the standard deviation of the intensity values.

\subsubsection{Architecture}

For this submission, 3D U-Net (Fig. \ref{fig:1st}-Top) is used as the main architecture which is based on nnU-Net implementation\footnote{https://github.com/MIC-DKFZ/nnUNet}. Compared to the original 3D U-Net, the notable changes \citep{isensee2019nnu} are the use of padding convolutions, instance normalization and leaky ReLUs. Predictions were made by three separate models and ensembled together with majority voting for kidney and an OR operation for tumor. One of these three models was a ``vanilla'' 3D U-Net, and the other two are cascaded models each with one network to localize the kidneys` Volumes of Interest (VOIs) and another to produce a fine-grained segmentation of the kidneys and tumors in each VOI.

At test time, the data was augmented with mirrors and predictions were averaged in an attempt to produce more robust predictions.


\subsubsection{Training}

The 3D U-Net model was first trained to minimize the sum of the cross-entropy and dice losses. For the cascaded models, the dice loss was replaced with a TopK loss to prevent tumor predictions in the upstream model. In all cases, the Adam algorithm was used for optimization with an initial learning rate of $3e-4$ followed by a fine-tuning learning rate of $3e-5$.

Each model took about four days to train with one Titan-Xp GPU and two Intel Xeon E5-2650V4 CPUs. 

\subsubsection{Postprocessing}

An heuristic-based algorithm was used to remove predicted kidney regions that were suspicious for false positives. First, the isolated points smaller then 20,000 voxels are removed. Second, the algorithm assumes that the centers of the two kidneys should have similar positions on the anterior-posterior axis, and larger isolated regions can be excluded if another plausible kidney region at that approximate position cannot be found.

\subsubsection{Results}

This submission scored a 0.973 kidney Dice and a 0.825 tumor Dice resulting in a fifth place 0.899 composite score. For a more detailed description of this submission, see \cite{kits19fifthplace}.

\section{Discussion}
\label{S:dis}

\subsection{Successful Segmentation Methods}

Semantic segmentation is one of the most popular research areas in medical image analysis, and as such, a vast number of novel sophisticated methods have been proposed in this space over the last few years. Surprisingly, very few of these are represented in the methods of the highest performing submissions, and this is consistent with the results of many concurrent challenges. The apparent failure of many of these methods to outperform e.g. the winning submission to this challenge is interesting and worthy of further study. Is this an artifact of this task in particular, or evidence of something more general? The winning method of Isensee et al. focused heavily on data preprocessing rather than novel architectures or optimization algorithms. Perhaps preprocessing has a larger impact than it typically gets credit for. Further experiments and challenges are needed to support or refute this claim.

\subsection{Prospects for Clinical Translation}

\revisednospace{}{Every case in the KiTS19 cohort included at least one renal tumor. Despite the usually unaffected contralateral kidney, algorithms exhibiting high performance on this segmentation task should not be trusted to provide the same level of performance on a cohort which includes healthy controls, and certainly not on the general population, for whom the incidence of renal tumors is very low. The intended use-case for algorithms developed based on this challenge is for patients who have already been diagnosed with a renal tumor by some means (possibly a second algorithm), and segmentation-based characterization of the tumor and kidney are desired (e.g., volume or surface area computation, enhancement pattern analysis, etc.).}

\revisednospace{}{Additionally, although this dataset contains image data from a diverse set of institutions and sites, a vast majority of this imaging was collected in Minnesota, North Dakota, and western Wisconsin. Further, assignment to training and test sets was fully random, making this an ``internal validation'' according to the risk of bias assessment outlined by the PROBAST instrument for evaluating multivariate clinical prediction models \citep{wolff2019probast}. Collecting a separate test dataset from a temporally and geographically disparate source would immensely improve the estimate of true generalization performance of these methods to future data and different centers around the world.}

\section{Limitations}
\label{S:lims}

The KiTS19 challenge was, overall, highly successful. It attracted a high number of submissions and continues to serve as an important and challenging benchmark in 3D segmentation. With that said, it was not perfect. In this section, we discuss limitations of this challenge, including issues with the dataset, issues with the challenge policies and design, and issues with the infrastructure for the challenge and some instances where communication between the organizers and participants broke down. We conclude with some ideas for how future iterations might address these limitations.

\subsection{Dataset}
\label{SS:ds}

Even though systems based on Deep Learning have often shown excellent generalization beyond the population that their training set was sampled from \citep{zhang2016understanding}, there is still reason for concern about a potential performance drop when applying these systems beyond the population that was sampled for the test set. 

The patients represented in the KiTS19 challenge were all treated by physicians within the same health system, and the population is therefore heavily concentrated in a limited geographic region, which might limit the generalizability of methods developed for this dataset to populations from other regions of the world. However, there is considerable diversity in imaging protocols and scanners since the preoperative studies were often performed at referring institutions. The dataset for the KiTS19 challenge was also retrospective, and the split into the training and test sets was random. Therefore, there is some concern that distributional shift over time might result in comparatively low performance on prospective data.

Even within this region and this block of time, the patients sampled represent only a subset of all patients seen for concern of renal malignancy. In particular, we were limited to patients who did not choose to ``opt out'' of making their data available for research, and we were limited to patients with contrast-enhanced imaging available for download (see Fig. \ref{fig:flow}). We have no reason to suspect that this introduced bias into our cohort, but this is, of course, difficult to check. Patients with tumor thrombus (27/329) were excluded in order to simplify the annotation process, and patients with concerning lesions that turned out to be cysts (2/329) were also excluded. The exclusion of cysts contributed to (but did not fully account for) the lower proportion of lesions postoperatively found to be benign in our cohort than has been reported elsewhere (8\% vs ~30\% -- \citep{kim2019association}). \revisednospace{Therefore, if a system based on this data were to be applied in clinical practice, care would need to be taken to ensure that the system was not being applied to patients who meet our exclusion criteria. Further, if a deployment`s target population differs significantly from that of this cohort (e.g., higher benign rate, lower tumor sizes -- see table \ref{tab:baselines}), the system might exhibit worse performance than on the KiTS19 test set.}{}

Finally, despite our best efforts to avoid errors when creating the semantic segmentation labels, they are imperfect. Errors range from inevitable noise in boundary delineations, to a few cases in which a whole structure was labeled incorrectly.

\subsection{Challenge Design}
\label{SS:des}

As stated in Section \ref{SSS:timeline}, when we first released the dataset, we designated a 20 day period for public ``label review'', in which participants were invited to visualize the segmentation labels and raise any concerns. While a number of important issues were discovered during this period, in retrospect we believe that this period should have been extended considerably, perhaps until just one month before the test set was released. \revised{This would have allowed for the few issues that were discovered after the data ``freeze'' date to be fixed for the MICCAI 2019 competition.}{Two issues\footnote{Some discussion about these issues can be found at \url{https://github.com/neheller/kits19/issues/21}} (on cases 15 and 37) were discovered after the data ``freeze'' and remained incorrect in the official training set for the the MICCAI 2019 competition. A later data freeze may have prevented this.}

Further, since we released the test images publicly, there is a possibility that teams may have manually intervened in the prediction process in order to illegally ``clean'' their predictions, or even simply segmented some cases manually from scratch. In order to mitigate this risk, we allowed only a two week period for submission, but we cannot exclude this possibility entirely. It might have been preferable to host our challenge on a platform where a kernel is made available with private access to the test data for prediction purposes only.

\revisednospace{}{Another limitation to this challenge is the relatively small proportion of submissions which made their solution open source. Some prior challenges have written a requirement for this into the rules of participation, and whether to do so for KiTS19 was the topic of considerable discussion among the challenge organizers. In the end, we decided to ``strongly encourage'' teams to release their code, but not require it. The primary reason we came to this decision was because we didn`t want the source code requirement to be a barrier to participation. The original goal of this challenge was to identify the best available methods for this segmentation problem, and we felt that the most effective way to do this was to attract as many teams as possible to participate. This strengthens the evidence that top-performing teams truly represent the state of the art.}

\revisednospace{}{In support of this point, several participants have indicated to us that a source code requirement would have prevented them from making their submission, including some of the highest scoring teams. As a concession, teams were asked to enclose short papers that included “sufficient detail for a third party to replicate your methods and reproduce your results” but we concede that some of the manuscripts that were submitted fall short of this standard, despite our repeated requests that they be revised and expanded. The most pronounced cases (6 of the original 106 submissions) resulted in disqualification.}

\revisednospace{}{There appears to be a tradeoff between the depth of reporting requirements that you can impose on participants and the quantity (and thereby quality) of top submissions the challenge will receive. In this challenge we prioritized the latter, but reasonable people might conclude that the former should take precedent.}

\revisednospace{}{In order to prevent cases of cheating via manual refinement, tt’s true that we could have privately collected each team’s source code and verified their predictions ourselves, but with 100 submitting teams using a wide range of frameworks and hardware configurations as well as potential dependencies on internal tooling, we decided that this would not be feasible with the resources available to us for this challenge. In the planned 2021 edition of KiTS, we will require submission of a Docker container such that inference is performed in a controlled environment and the test imaging never leaves our institution. We suspect that this will help to encourage teams to release their code since they will be sending it to us anyways, but at present we still plan for this to be optional since we highly value contributions from industry-affiliated teams, even though they may have insurmountable barriers to releasing code to the public.}

\subsection{Communication and Challenge Infrastructure}
\label{SS:inf}

There were two public avenues for communication between the participants and organizers: GitHub issues and a Discourse forum. GitHub Issues were meant to be used for personal issues with downloading and using the data, as well as for reporting label errors, where Discourse was to be used for everything else. The intention was that Discourse would be for information that benefits all participants and GitHub was for troubleshooting individual difficulties. In retrospect, the reporting of label errors was wrongly confined to GitHub, and was therefore not widely disseminated. In future challenges, we will take conscious steps to ensure that all participants are made aware of issues that are found with the training labels so that teams are given equal opportunities to exclude or amend these issues for their training process.

\subsection{Future Directions for KiTS}
\label{SS:future}

We are actively working to improve upon the KiTS19 Challenge in several ways. Among these are:

\begin{itemize}
    \item \textbf{Multi-Institutional Cohort:} We will be expanding the dataset to represent at least four health systems, each in different geographic regions.
    \item \textbf{Pseudo-Prospective Cohort:} A date will be selected, and data generated before that date will be used only for training, and data generated after that time will be used only for testing. 
    \item \textbf{Longer Data Review:} We will be extending the time period in which concerns will be addressed prior to the data freeze. Tentatively, we plan to freeze the data just one month before the test set release.
    \item \textbf{Clearer Communication of Label Errors:} In addition to the labels` version control system, label errors discovered after the data freeze will be announced on the homepage of the challenge as well as on the discussion forum.
    \item \textbf{Better Representation of Rare Subtypes:} A vast majority of renal tumors are Clear Cell Renal Cell Carcinomas, and this is reflected in the KiTS19 dataset. With contributions from other clinical centers, we will have enough data to perform stratified random sampling in order to give equal representation to several histological subtypes that are comparatively rare.
    \item \textbf{More Segmentation Classes:} In order to prevent the segmentation problem from becoming trivially easy with the larger dataset, we plan to expand the segmentation problem to include more classes and structures such as renal cyst, renal artery and vein, and ureter.
\end{itemize}

Things such as a longer data review and clearer communication of label errors are simple to implement, but others such as more segmentation classes and multi-institutional data will take significant effort. Our hope is to phase these changes into future KiTS Challenges as time and administrative hurdles allow.

\section{Conclusion}
\label{S:conc}
The KiTS19 challenge served to accelerate and measure the state of the art in the automatic semantic segmentation of kidneys and kidney tumors in contrast-enhanced CT imaging. The challenge attracted submissions from more than 100 teams around the world, and the highest-scoring team achieved a kidney Dice score of 0.974 and a tumor Dice score of 0.851 on the private 90-case test set. The experiments and results of the winning team are surprising in that they failed to show any meaningful benefit to several ``bells and whistles'' that people have recently reported to yield substantial improvements over the 3D U-Net baseline. Instead, they won by a considerable margin by submitting the predictions of the baseline+residual connections model alone. The challenge has now entered an indefinite ``open leaderboard'' phase where it serves as a high-quality and challenging benchmark in 3D semantic segmentation. A second iteration of the KiTS challenge is planned with the goals of improving upon the clinical significance and external validity of the challenge, as well as increasing the difficulty of the 3D segmentation problem by adding other more complicated structures such as ureters, renal arteries, and renal veins.

\section*{Acknowledgments}
Research reported in this publication was supported by the National Cancer Institute of the National Institutes of Health under Award Number R01CA225435. The content is solely the responsibility of the authors and does not necessarily represent the official views of the National Institutes of Health.

We would like to thank the MICCAI challenge committee for taking the time to review our challenge proposal and provide useful feedback. We also thank grand-challenge.org for providing an excellent free platform for hosting challenges such as this one. Finally, we thank the developers of Discourse for providing an excellent piece of free software for self-hosted discussion forums.

\bibliographystyle{model2-names.bst}\biboptions{authoryear}
\bibliography{main.bbl}


\end{document}